\documentclass[aps,prl,twocolumn,notitlepage,showpacs,superscriptaddress,am]{revtex4-2}%
\usepackage{graphicx}
\usepackage{amsmath}
\usepackage{amssymb}
\usepackage{color}
\usepackage{amsfonts}%
\providecommand{\U}[1]{\protect\rule{.1in}{.1in}}

\def\Herb{$\rm ZnCu_3(OH)_6Cl_2$}
\def\GaHerb{$\rm GaCu_3(OH)_6Cl_3$}
\def\YHerb{$\rm YCu_3(OH)_6Cl_3$}
\def\Ykap{$\rm Y_3Cu_9(OH)_{18}OCl_8$}
\def\Y{Y-kapellasite}

\def\aI3{$\alpha$-I$_3$}

\def\iTone{$T_1^{-1}$}

\def\sro{Sr$_2$RuO$_4$}

\begin{document}
\title{Controlling frustrated magnetism on the kagome lattice by uniaxial-strain tuning}

\author{Jierong Wang}
\affiliation{Department of Physics and Astronomy, UCLA, Los Angeles, California 90095, USA}
\author{Y.-S. Su}
\affiliation{Department of Physics and Astronomy, UCLA, Los Angeles, California 90095, USA}
\author{M. Spitaler}
\affiliation{Institute of Solid State Physics, TU Wien, 1040 Vienna, Austria}
\author{K.M. Zoch}
\affiliation{Institute of Physics, Goethe-University Frankfurt, 60438 Frankfurt (Main), Germany}
\author{C. Krellner}
\affiliation{Institute of Physics, Goethe-University Frankfurt, 60438 Frankfurt (Main), Germany}
\author{P. Puphal}
\affiliation{Institute of Physics, Goethe-University Frankfurt, 60438 Frankfurt (Main), Germany}
\affiliation{Max-Planck-Institute for Solid State Research, 70569 Stuttgart, Germany}
\author{S. E. Brown}
\affiliation{Department of Physics and Astronomy, UCLA, Los Angeles, California 90095, USA}
\author{A. Pustogow}
\email{pustogow@ifp.tuwien.ac.at}
\affiliation{Department of Physics and Astronomy, UCLA, Los Angeles, California 90095, USA}
\affiliation{Institute of Solid State Physics, TU Wien, 1040 Vienna, Austria}


\begin{abstract}
It is predicted that strongly interacting spins on a frustrated lattice may lead to a quantum disordered ground state or even form a quantum spin liquid with exotic low-energy excitations. However, a thorough tuning of the frustration strength, separating its effects from those of disorder and other factors, is pending.
Here we break the symmetry of a kagome-lattice compound in a controlled manner by applying \textit{in situ} uniaxial stress. The transition temperature of \Ykap\ is linearly enhanced with strain, $\Delta T_{\rm N}/T_{\rm N} \approx 10\%$ upon in-plane compression of order 1\%, providing clear evidence for a release of frustration and its pivotal role for magnetic order. 
Our comprehensive $^1$H NMR results suggest a $\overrightarrow{Q}=(1/3\times 1/3)$ state under unstrained conditions and further reveal an incomplete antiferromagnetic transition with fluctuating moments in this strongly frustrated system.  
\end{abstract}

\maketitle

\begin{figure}[b]
\centering
\includegraphics[width=1\columnwidth]{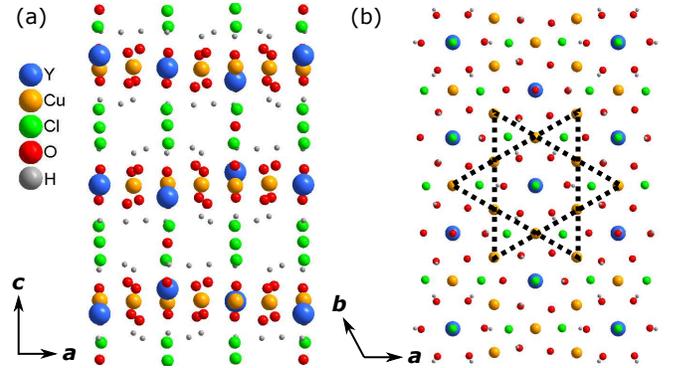}
\caption{Crystal structure of \Ykap\ (\Y ). (a) Cu$^{2+}$ atoms (orange) are arranged in layers parallel to the $ab$-plane. (b) Within the plane they form a $S=1/2$ kagome lattice indicated by black dotted lines.
}
\label{structure}
\end{figure}

Even after two decades of intense scrutiny, quantum spin liquids (QSL) remain an elusive state of matter~\cite{Balents2010,Savary2017,Zhou2017,Broholm2020}. Apart from growing evidence for the importance of disorder \cite{Olariu2008,Freedman2010,Zhu2017,Li2017,Kimchi2018,Itou2017,Riedl2019,Pustogow2020HgCl,Smaha2020a, Miksch2021,Pustogow2022}, geometrical frustration is considered decisive to suppress magnetic order in presence of strong antiferromagnetic (AFM) exchange interactions. The vast majority of QSL candidates are found in quasi two-dimensional correlated electron systems with triangular \cite{Shimizu2003,Isono2014,Shimizu2016,Itou2010,Shen2016,Xu2016,Li2017}, honeycomb \cite{Takagi2019} 
or kagome \cite{Puphal2018,Norman2016} lattices. 
Herbertsmithite, \Herb, is an archetype realization of the latter symmetry \cite{Norman2016} and has been intensely studied over the last two decades~\cite{Braithwaite2004,Shores2005,Olariu2008,Han2012,Fu2015,Norman2016,Zorko2017,Khuntia2020,Wang2021} 
-- not least due to the exciting proposal of exotic superconductivity and Dirac bands in a doped kagome lattice \cite{Mazin2014}. Although the latter scenario could not be realized so far \cite{Puphal2019}, many related compounds substituting Zn by other bi- or trivalent cations have been synthesized by now~ \cite{Puphal2018}. Among those, \Ykap\ \cite{Puphal2017} (denoted as \Y ) and \YHerb\ \cite{Sun2016}  crystallize in the closely-related kapellasite structure (see Fig.~\ref{structure}) and exhibit AFM order at temperatures $T_{\rm N}\ll J/k_{\rm B} \approx 10^2$~K.

So far, most attempts to modify the frustration strength focused on chemical substitution in order to arrange the valence electrons in the above mentioned patterns. While commonly physical pressure is applied to tune electronic interactions, e.g. towards metal-insulator transitions, hydrostatic compression does not directly affect the lattice symmetry, unless it triggers a structural transition. Recent developments in piezoelectric uniaxial strain applications at cryogenic temperatures \cite{Hicks2014a,Barber2019} 
 now provide us the opportunity to modify the degree of geometrical frustration in a controlled manner. 
 
Here, we take full advantage of uniaxial strain to directly tune magnetic order in \Y\ single crystals.
We characterize the magnetic properties by $^1$H nuclear magnetic resonance (NMR) in a temperature range 1.5--200~K and reveal strong spin correlations for $T<30$~K. We find AFM below $T_{\rm N}=2.2$~K that is consistent with the proposed $\overrightarrow{Q}=(1/3\times 1/3)$ order~\cite{Hering2022}. By applying uniaxial strain of order 1\% we tune the exchange interactions and frustration strength \textit{in situ} triggering a pronounced increase of $T_{\rm N}$ linear with strain.

\begin{figure*}[ptb]
\centering
\includegraphics[width=2.05\columnwidth]{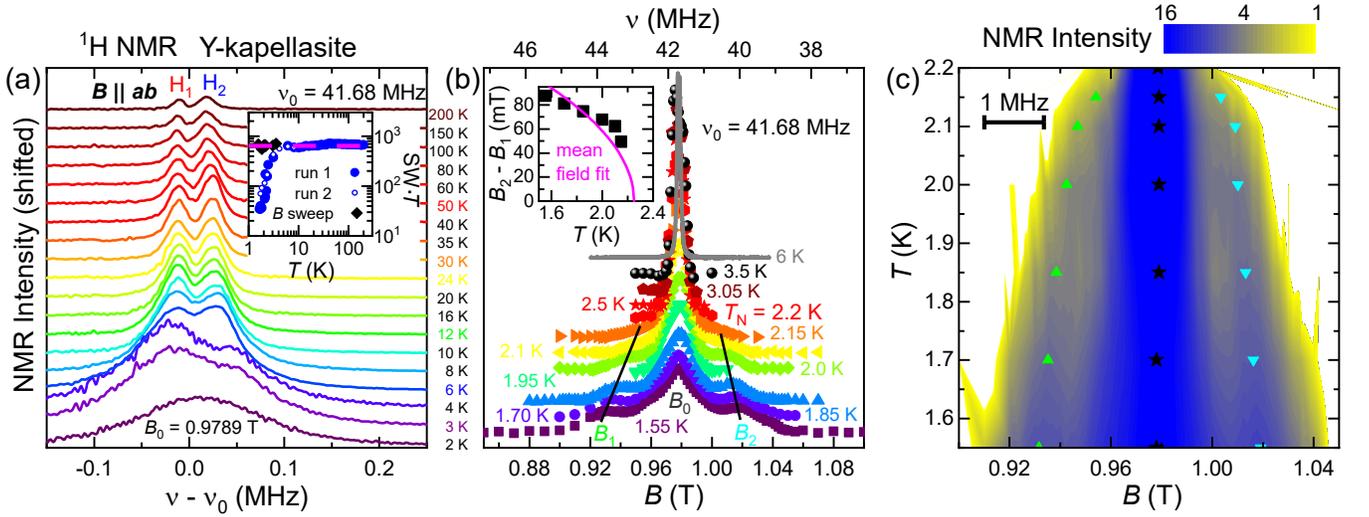}
\caption{(a) The $^1$H NMR spectra acquired for in-plane magnetic field ($B\perp a$, $B_0=0.9789$~T) consist of two peaks ($H_1$, $H_2$) separated by 30--40 kHz. The lines exhibit successive broadening upon cooling and exceed the experimental bandwidth below 4~K, as seen by the deviation of the integrated spectral weight $SW$ from the standard $T^{-1}$ dependence (inset). (b) The spectra below 4~K, acquired by magnetic field sweeps at constant frequency $\nu_0=41.68$~MHz, reveal a pronounced  splitting into three peaks below $T_{\rm N}=2.2$~K. The separation between the two outer peaks ($B_1$, $B_2$) increases to lower temperatures (inset) while the central peak remains at $B_0$. The grey line indicates the spectrum at 6~K, which follows the frequency scale at the top.  (c) The false-color plot of the data from panel (b) illustrates the shifting of $B_1$ and $B_2$ as well as the strong broadening. The horizontal bar corresponds to a frequency of $\Delta \nu = \Delta B/^1\gamma=1$~MHz.
}
\label{spectra}
\end{figure*}
In the two Y analogs of herbertsmithite (kapellasite), the additional charge upon substitution of Zn$^{2+}$ is compensated, resulting in Mott insulators with a charge-transfer gap of 3~eV ($U\approx 8$~eV \cite{Pustogow2017Herbertsmithite}). In the case of \YHerb\ this leads to an unstable crystal structure -- similar to \GaHerb\ \cite{Puphal2019} it can be only synthesized as powder \cite{Sun2016,Barthelemy2019} -- while large \Ykap\ single crystals with slightly distorted kagome layers (structure shown in Fig.~\ref{structure}) can be grown by hydrothermal methods \cite{Puphal2017}.  
Both compounds exhibit AFM order at temperatures much lower than $\Theta_{CW}\approx 100$~K; magnetization, specific heat, $\mu$-SR and neutron diffraction experiments yield $T_{\rm N}=2.2$~K for \Y\ \cite{Puphal2017,Barthelemy2019}, related with the observation of THz magnons~\cite{Biesner2022}, and $T_{\rm N}=15$~K for \YHerb\ \cite{Sun2016,Barthelemy2019,Zorko2019,Zorko2019a}. 
Despite the apparent absence of a QSL state down to $T\rightarrow 0$, distorted kagome lattices~\cite{Ishikawa2015,Boldrin2015,Boldrin2016,Spachmann2022} came into focus recently due to magnetoelastic coupling and the realization of non-trivial ground states such as pinwheel~\cite{Matan2010,Smaha2020} and $\overrightarrow{Q}=(1/3\times 1/3)$ order~\cite{Hering2022}.

Prior to application of strain we carried out a thorough ambient-pressure NMR characterization of \Y. The magnetic field was aligned parallel to the Cu$^{2+}$ chains of the kagome network, i.e. perpendicular to the crystallographic $a$-axis; in the following, this in-plane field configuration is denoted as $B\parallel ab$. Fig.~\ref{spectra}a presents the temperature evolution of the $^1$H spectra ($B_0=0.9789$~T, $\nu_0=41.68$~MHz), yielding two peaks separated by 30--40~kHz above 4~K. At lower temperatures, the line broadening exceeds the experimental bandwidth such that the intensity (probed by integrating the spectral weight $SW$; shown in inset of Fig.~\ref{spectra}a) deviates from the standard $T^{-1}$ dependence. To resolve the entire peak structure, we performed magnetic field sweeps at constant frequency $\nu_0$ covering the temperature range between 1.55~K and 3.5~K (Fig.~\ref{spectra}b,c). Below $T_{\rm N}$ the signal splits into a triplet: while the central line ($B_0$) remains unshifted, the two outer peaks ($B_1$, $B_2$) move apart symmetrically upon cooling. As plotted in the inset, the separation  $B_2-B_1$ is consistent with the onset of a broken symmetry, which is further illustrated by the yellow-blue contour plot in Fig.~\ref{spectra}c. On a quantitative level, the line splitting at the lowest measured temperature agrees well with a local field $B_{loc}\approx 60-64$~mT originating from dipolar coupling between $^1$H nuclear moments and AFM ordered electron spins on the Cu$^{2+}$ sites (see Fig.~\ref{structure}), which are at a distance 2.44--2.50~\AA\ \cite{Puphal2017}. 

\begin{figure}[ptb]
\centering
\includegraphics[width=1\columnwidth]{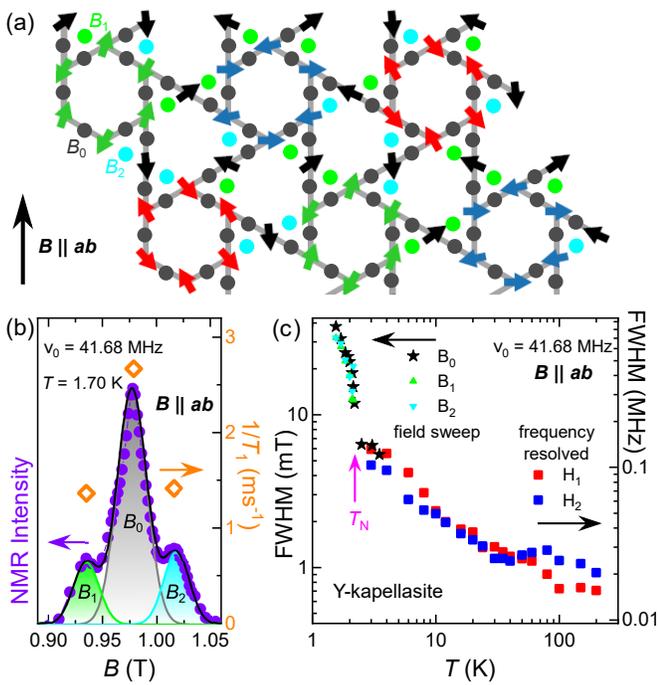}
\caption{(a) $Q=(1/3\times 1/3)$ AFM order predicted for a distorted kagome lattice~\cite{Hering2022}. Colored circles indicate location of $^1$H nuclei in the crystal lattice of \Y\ (cf. Fig.~\ref{structure}). (b) The NMR intensity (violet) at 1.7~K was fitted with three Gaussian peaks corresponding to the $^1$H positions (see color) in (a); their intensity scales as 4:0.93:1.07 ($B_0$:$B_1$:$B_2$).  The spin-lattice relaxation rate (orange diamonds) was measured at different fields: \iTone\ of the side peaks ($B_1$, $B_2$) is  $\approx 50$\% smaller than for the central peak at $B_0=0.9789$~T. (c) The full width at half maximum (FWHM), extracted for the $^1$H NMR peaks $H_1$ and $H_2$ ($T>T_{\rm N}$; see Fig.~\ref{spectra}a) as well as for the three lines $B_0$, $B_1$ and $B_2$ emerging below $T_{\rm N}$ (see Fig.~\ref{spectra}b,c), exceeds the homogeneous linewidth below 30~K and shows a steep increase at $T_{\rm N}$. Note the different units of the vertical scales. 
}
\label{T1-linewidth}
\end{figure}

While our observation of a line splitting evidences AFM order setting in at $T_{\rm N}$, the ratio among the peak intensities provides insight into the details of the spin structure. Upon $\overrightarrow{Q}=(1/3\times 1/3)$ order predicted for a distorted kagome lattice~\cite{Hering2022}, along a Cu$^{2+}$ chain four successive AFM bonds are satisfied followed by two unsatisfied bonds. We point out that $^1$H nuclei are ideal probes of the local field at a bond as they are attached to oxygen atoms and, hence, are sitting between two neighboring Cu$^{2+}$ sites, as sketched in Fig.~\ref{T1-linewidth}a. If the electron spins are antiparallel, their local fields cancel out yielding a proton resonance at $B_0$; for parallel alignment, their local fields add up to (subtract from) the external field yielding the NMR line at $B_1$ ($B_2$). 
Our Gaussian fits in Fig.~\ref{T1-linewidth}b indeed reveal an intensity ratio between $B_0$:$B_1$:$B_2$ sites of 4:0.93:1.07 that is close to 
4:1:1~\footnote{Strictly speaking, the addition/subtraction of local fields depends on the angle to the external field, which acquires various values between [$0^{\circ}$,$180^{\circ}$]. Also, there are several inequivalent $^1$H positions in the crystal structure of \Y. Nevertheless, we expect distributions peaked at specific field values that are symmetric around $B_0$. Dedicated calculations based on the $\overrightarrow{Q}=(1/3\times 1/3)$ model \cite{Hering2022} are desired in future work. }. Overall, the spectral features of our NMR measurements are consistent with $\overrightarrow{Q}=(1/3\times 1/3)$ order in \Y\ \cite{Hering2022} and motivate more thorough assessment of this model.

Apart from the splitting, the NMR linewidth (Fig.~\ref{T1-linewidth}c) is subject to severe broadening. The full width at half maximum (FWHM) increases beyond the homogeneous linewidth $(\pi T_2)^{-1}\approx 35$~kHz already for $T<30$~K, indicating the onset of antiferromagnetic fluctuations, followed by an abrupt broadening at $T_{\rm N}$. Note, while the FWHM increases by two orders of magnitude across the studied temperature range, the spin-spin relaxation rate merely varies by a factor two. The two-peak structure of the spectrum ($T>4$~K) and the absolute values of $T_2$ are caused by proton-proton dipolar coupling since hyperfine coupling to the spin-1/2 Cu$^{2+}$ sites is weak. Hence, we assign the pronounced line broadening below 30~K to magnetic correlations in \Y.

\begin{figure}[ptb]
\centering
\includegraphics[width=1\columnwidth]{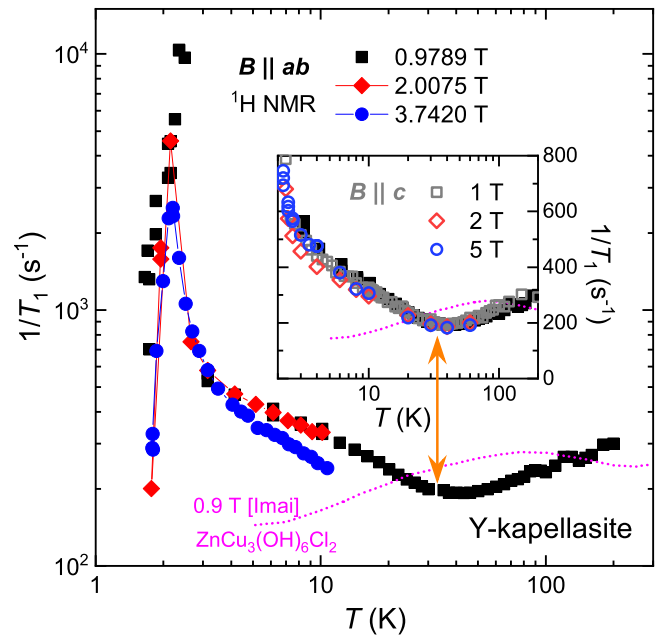}
\caption{$^1$H NMR spin-lattice relaxation rate \iTone\ for $B\parallel ab$ and $B\parallel c$ (inset). Upon cooling from room temperature, \iTone\ initially decreases towards a minimum around 32 K (orange double arrow). Below that, the increase of \iTone\ signals the onset of strong antiferromagnetic spin fluctuations, which result in a sharp maximum at $T_{\rm N}=2.2$~K. Increasing the magnetic field yields a reduced peak value and broadening of the transition, consistent with specific heat results \cite{Puphal2017}. Above 4~K \iTone\ does not exhibit pronounced field dependence or anisotropy. 
}
\label{T1-ambient}
\end{figure}

Aside from $T_2$ effects, the spin-lattice relaxation rate \iTone\ is highly susceptible to the emerging magnetism. 
Above 10~K the $^1$H relaxation rate of \Y, plotted in Fig.~\ref{T1-ambient}, is of similar magnitude compared to \Herb\ \cite{Imai2008}. However, herbertsmithite exhibits a different temperature dependence with a maximum around 90~K and steadily decreasing \iTone\ upon cooling below that (magenta dotted line in Fig.~\ref{T1-ambient}). Here, we observe a positive slope at high temperatures, likely related to lattice dynamics, while below 30~K the relaxation rate increases upon cooling due to AFM fluctuations. Down to 4~K we do not observe pronounced field dependence or anisotropy (measurements for $B\parallel c$ shown in inset of Fig.~\ref{T1-ambient}) in \iTone.

The onset of magnetic order yields a sharp peak at the transition temperature $T_{\rm N}=2.2$~K.
Increasing the magnetic field to 2~T and 3.7~T decreases the peak values of \iTone\ and broadens the maximum, in accord with specific heat results \cite{Puphal2017}. As AFM sets in, relaxation rapidly decreases upon further cooling. Throughout, we determined the temperature dependence of \iTone\ at $B_0$. At 1.70~K we also measured \iTone\ at the two outer peaks $B_1$ and $B_2$ (Fig.~\ref{T1-linewidth}b) yielding a 50\% smaller value compared to $B_0$. The faster relaxation at the central peak indicates fluctuating spins -- possibly a small portion that has evaded (or is not involved in) static AFM order. In this regard, NMR measurements at lower temperatures are highly desired, especially at $T\ll 1$~K.

\begin{figure}[ptb]
\centering
\includegraphics[width=1\columnwidth]{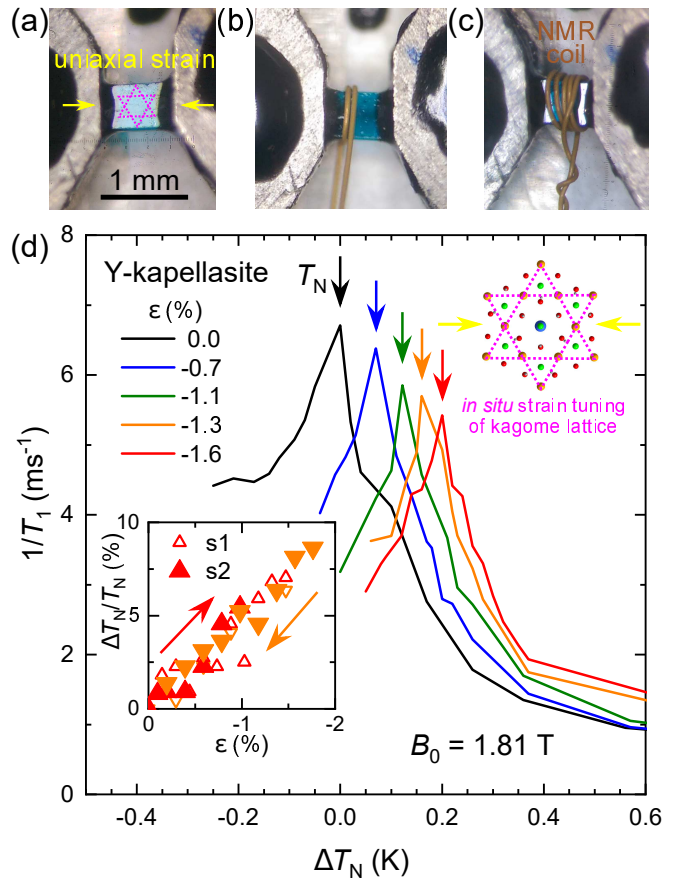}
\caption{(a-c) For NMR experiments under uniaxial strain a single crystal was glued between the two arms of a piezoelectric strain cell (a) and, subsequently, an NMR coil was wound around it (b,c). 
(d) \iTone\ was measured for $B\parallel a$ upon uniaxial compression of the kagome lattice parallel to the Cu$^{2+}$ chains (sketched on top right; cf. Fig.~\ref{structure}). Inset: in-plane strain of $1.5\%$ results in a considerable enhancement of the transition temperature by $\Delta T_{\rm N}/T_{\rm N}\approx 10\%$. 
}
\label{strain}
\end{figure}

To summarize the findings so far, the NMR properties are consistent with AFM order of $\overrightarrow{Q}=(1/3\times 1/3)$ symmetry~\cite{Hering2022}. Moreover, \Y\ features coexistence of fluctuating and static moments, in line with recent $\mu$-SR results~\cite{Barthelemy2019}. 
The spin-lattice relaxation rate is highly susceptible to the onset of magnetic order and exhibits a sharp peak at the transition, making it a sensitive probe of the transition temperature. Having identified its unusual magnetic ground state, in the following we apply in-plane uniaxial stress to \Y\ single crystals. Through distorting the kagome lattice by compression along the Cu$^{2+}$ chains, as sketched in Fig.~\ref{strain}d, we directly modify the anisotropy of the transfer integrals $t'/t$ and, hence, the exchange interactions $J$, $J'$ and their degree of geometrical frustration.

In Fig.~\ref{strain} we trace the change in transition temperature through \iTone\ measurements at $B_0=1.81$~T ($B\parallel a$) upon applying uniaxial stress parallel to the Cu$^{2+}$ chains. There is a clear enhancement of $T_{\rm N}$ with approximately linear dependence on the applied strain, which agrees with calculations of Heisenberg AFM ordering on a distorted kagome lattice~\cite{Masuda2012}. Our results on two samples consistently reveal an increase of transition temperature by almost 10\% for compressive strain of order 1\% parallel to the kagome layers. How does this compare with the modifications expected for hydrostatic pressure of similar size? For comparison, we consider that the superexchange is proportional to $t^4/(\Delta^2 U)$ and $t\propto r^{-4}$, hence 1\% reduction of Cu-Cu distance should result in an increase of 16\%. 
However, this is a large overestimate, because in real materials the crystal lattice adapts mostly by changing bond angles rather than bond length~\footnote{A notable exception is \sro\ where the length changes upon uniaxial strain do not involve a change of bond angles \cite{Steppke2017,Luo2019}.}. Thus, we conclude that the major effect arises from releasing frustration of the kagome lattice. 

Let us assess the observed strain-tuning effects in the context of other materials where the structure impacts the magnetic properties.
In \Herb, indications of DM interaction \cite{Norman2016,Zorko2017} and even symmetry breaking \cite{Laurita2019,Norman2020} have been reported, likely related to magneto-elastic coupling \cite{Li2020}. Hydrostatic pressure yields an emergence of magnetic order \cite{Kozlenko2012}, but also a field-induced spin freezing was reported \cite{Jeong2011}. The compound studied here shows structural similarities to the 'pin-wheel' kagome structure of Rb$_2$Cu$_3$SnF$_{12}$~\cite{Matan2010,Grbic2013} and a Barlowite polymorph~\cite{Smaha2020}. To that end, \Y\ is likely frustrated in a different way than Herbertsmithite or Kapellasite, which both exhibit an undistorted kagome lattice -- at the expense of severe Zn/Cu antisite disorder \cite{Freedman2010}. Finally, an indication for a structural involvement in the magnetic degrees of freedom is the onset of spin fluctuations below a recently discovered structural anomaly at 32~K~\cite{Chatterjee2022}. An intriguing scenario could be that this is the onset of nematic order below a crossover at $T^{\star}\gg T_{\rm N}$ predicted for a distorted kagome lattice~\cite{Masuda2012}. Extending our present strain-tuning experiments on \Y\ to higher temperatures would enable direct scrutiny of this issue and thus remains a desideratum for future work.

To conclude, we performed comprehensive $^1$H NMR investigations on \Y\ as a function of temperature, magnetic field and uniaxial stress. Under unstrained conditions we find spectral evidence for AFM order of $\overrightarrow{Q}=(1/3\times 1/3)$ type below $T_{\rm N}=2.2$~K in this distorted kagome compound. 
Clear signatures for AFM are also seen in the linewidth and spin-lattice relaxation rate, which are dominated by proton dynamics at elevated temperatures while magnetic correlations become dominant below 30~K. 
Based on this characterization, we applied \textit{in situ} uniaxial stress parallel to the kagome layers, resulting in a linear increase of $T_{\rm N}$ with strain, in accord with theoretical predictions~\cite{Masuda2012}. Through a controlled release of frustration, our findings evidence its crucial importance for correlated quantum magnets and spin liquids, opening the door for similar studies in triangular, honeycomb and other kagome materials. The pioneering strain-tuning approach presented here is ideally suited to resolve the magnetic ground states of existing materials, and to discover novel exotic spin phases beyond state-of-the-art theoretical frameworks \cite{Savary2017,Broholm2020}.

\section*{Methods}

\textbf{NMR experiments} were performed using variable temperature $^4$He cryostats and superconducting magnets. At $T\geq 4$~K magnetic field and frequency were fixed according to the $^1$H resonance condition $\nu_0=^1\gamma B$, with $^1\gamma=42.5774$~MHz/T, as the proton NMR spectrum was resolved within the experimental bandwidth. Due to excessive line broadening below 4~K, the NMR intensity started to drop below the expected $T^{-1}$ behavior, as illustrated by blue circles in the inset of Fig.~\ref{spectra}a that correspond to the integrated spectral weight ($SW$) of the NMR spectrum.
To that end, at $T<4$~K the $^1$H spectra were acquired by sweeping the external field in dense intervals around $B_0=0.9789$~T. As illustrated by the black diamonds in the inset of Fig.~\ref{spectra}a, the entire NMR intensity was recovered in these field-sweep measurements: in the inspected field range the total $SW$ of the $^1$H resonance lines up with the $T^{-1}$ behavior extrapolated from $T\geq 4$~K.
Zero-strain NMR measurements were performed using large single crystals ($3\times 2\times 1$~mm$^3$). The relaxation measurements (Fig.~\ref{T1-ambient}) for in-plane (measured at UCLA) and out-of plane (acquired at TU Wien on a different sample) crystal axes show similar \iTone\ values. The relaxation rate \iTone\ was acquired at the central line ($\nu_0$, $B_0$ in Fig.~\ref{spectra}b). At $T=1.70$~K we measured \iTone\ at $\nu_0=41.68$~MHz also for the two satellite peaks $B_1$ and $B_2$ obtaining values approximately 50\% of \iTone/ at $B_0=0.9789$~T (orange symbols in Fig.~\ref{T1-linewidth}b).

\textbf{Uniaxial strain experiments} were performed using piezoelectric strain cells ($Razorbill$) in the same $^4$He cryostat as the zero-strain studies for in-plane magnetic fields. For that, the \Y\ sample (the single crystal shown in Fig.~\ref{strain}a-c was cut in dimensions $2.5\times 0.45\times 0.24$~mm$^3$ with the longest dimension parallel to the Cu chains of the kagome layer, i.e. $\varepsilon\perp a$) was glued at room temperature on the strain cell using Stycast and dried for several days. Here, the magnetic field direction was parallel to the $a$-axis, i.e. perpendicular to the in-plane fields in Fig.~\ref{spectra}. Still, we observed a similar peak in \iTone\ at $T_{\rm N}$ for both $B\parallel a$ (Fig.~\ref{strain}) and $B\parallel ab$ (Fig.~\ref{T1-ambient}), 
hence our measurements of the relaxation rate were suitable to trace the strain dependence of the AFM transition.
The strain voltage was varied at 4~K and successively \iTone\ was measured upon cooling through the transition, both for increasing and decreasing strain as indicated by red and orange symbols in Fig.~\ref{strain}d, respectively. The compression was monitored \textit{in situ} by the built-in capacitive displacement sensor. Similar to previous strain studies~\cite{Luo2019,Chronister2022}, the applied strain $\varepsilon$ was calculated as the displacement divided by the length of the sample surface that remained free of epoxy -- for sample s2 this corresponds to 0.6~mm (Fig.~\ref{strain}a-c).  
The linear increase of $T_{\rm N}$ with strain shown in the inset of Fig.~\ref{strain}d lines up for the two samples. Still, we do not exclude that the $real$ compression of the sample is somewhat lower than the present estimate using the built-in capacitive sensor, likely $\varepsilon_{real}\approx 1$\%; also in \sro\ the strain of the van Hove singularity was first estimated as 0.6\% \cite{Steppke2017,Luo2019} and later corrected to 0.44\% by using high-precision stress-strain sensors \cite{Barber2019a}.

\acknowledgments We thank M. R. Norman, M. Dressel, H. Jeschke and  I. I. Mazin for useful comments and discussions. Support with sample preparation by G. Untereiner is kindly appreciated. A. P. acknowledges support by the Alexander von Humboldt Foundation through the Feodor Lynen Fellowship. Work at UCLA was supported by the National Science Foundation (DMR-2004553).


%

\end{document}